\documentclass[article,scriptaddress,twocolumn, prr,showkeys,showpacs,superscriptaddress,longbibliography]{revtex4-1}

\usepackage{amsmath,amssymb} 
\usepackage{bm}
\usepackage{esvect}
\usepackage{graphicx, xcolor}  
\usepackage{dcolumn}   
\usepackage{pifont}

\begin{document}

\title{Migration feedback induces emergent ecotypes and abrupt transitions in evolving populations}

\author{Casey O. Barkan}
\author{Shenshen Wang}
\affiliation{Department of Physics and Astronomy, University of California, Los Angeles, Los Angeles, CA 90095}

\date{\today}

\begin{abstract}

We explore the connection between migration patterns and emergent behaviors of evolving populations in spatially heterogeneous environments. 
Despite extensive studies in ecologically and medically important systems, a unifying framework that clarifies this connection and makes concrete predictions remains much needed.
Using a simple evolutionary model on a network of interconnected habitats with distinct fitness landscapes, we demonstrate a fundamental connection between migration feedback, emergent ecotypes, and an unusual form of discontinuous critical transition. We show how migration feedback generates spatially non-local niches in which emergent ecotypes can specialize. Rugged fitness landscapes lead to a complex, yet understandable, phase diagram in which different ecotypes coexist under different migration patterns. The discontinuous transitions are distinct from the standard first-order phase transitions in statistical physics. They arise due to simultaneous transcritical bifurcations and exhibit a ``fine structure" due to symmetry breaking between intra- and inter-ecotype interactions. We suggest feasible experiments to test our predictions.

\end{abstract}

\maketitle

\section{Introduction}

Biological populations live and evolve in spatially heterogeneous environments, where selection pressures depend on location. While certain consequences of spatial heterogeneity are well established -- e.g. that heterogeneity can stabilize biodiversity \cite{rainey1998adaptive, chesson2000mechanisms, amarasekare2003competitive, dias1996sources, hermsen2010sources, hermsen2012rapidity, roy2020complex} -- recent studies have uncovered unexpected behaviors, such as persistent fluctuations in species abundance \cite{hu2022emergent} and chaotic alternation of locally dominant strains \cite{pearce2020stabilization}, non-monotonic effects of dispersal ability \cite{lombardo2014nonmonotonic, de2018tuning, barkan2023multiple}, and re-entrant transitions \cite{barkan2023multiple}. These findings highlight the need for a more comprehensive understanding of how migration patterns influence populations in spatially-structured environments.

Spatial structure in medically important systems produces complex behaviors and raises important questions. Human gut microbiota inhabit several distinct regions with differing physiological conditions (e.g. nutrient concentrations \cite{thursby2017introduction}) and migration of microbes is modulated by active mechanisms (e.g. contraction of the colonic walls  \cite{arnoldini2018bacterial}). Flow speed and pattern were found to strongly shape the spatiotemporal composition of the microbiome \cite{cremer2016effect, cremer2017effect}. Infectious bacteria exposed to antibiotics live within, and transfer between, distinct hosts and clinical settings \cite{krieger2020population}, and their migration patterns can affect evolution of antibiotic resistance \cite{fraebel2017environment, gude2020bacterial,shaer2022multistep}. 
In metastatic cancers, invasion of new tissue sites and an increased diversity of motility phenotype are often associated with the emergence of drug resistance and systemic failure \cite{housman2014drug}.
In the adaptive immune system, antibody-producing B-cells undergo rapid evolution within localized germinal centers \cite{victora2012germinal}. It remains an open question to what extent the strength and pattern of migration between germinal centers shape clonal diversity and response efficacy. 

A general approach for modeling spatially heterogeneous systems is to treat the environment as a network of interconnected habitats, where each habitat has distinct selection pressures. There is a long history of such models in the ecology and evolution literature, which has revealed how migration rates can control the spatial scale of adaptation \cite{bodmer1968migration, felsenstein1976theoretical,kawecki2004conceptual,economo2008species,spichtig2004maintenance}. In particular, low migration leads to spatially localized adaptation (where evolution favors organisms specialized to their local environment), whereas high migration rates promote \textit{generalist} organisms that are globally well-adapted. However, previous studies assume neutral evolution or simple unimodal fitness landscapes in each habitat, whereas realistic fitness landscapes may well be rugged \cite{palmer2015delayed, papkou2023rugged}. In addition, interesting results on the influence of migration patterns have been found \cite{tufto1996inferring}, yet a general understanding of the impact of different forms of migration patterns is lacking. Furthermore, although many of these models exhibit transitions between regimes \cite{kisdi2002dispersal,hanski2011eco,brown1992evolution,spichtig2004maintenance}, the exact nature of these transitions is often unclear.

In this paper, we study a model of an evolving population on a network of habitats, and uncover general principles by which migration patterns shape the population composition and critical transitions. We find that migration feedback (i.e. `loops' in migration patterns) allow for the evolution of emergent ecotypes that adapt to fill spatially non-local niches. At high migration rates, spatial coexistence of distinct ecotypes is lost, and evolution leads to a generalist only if there is migration feedback. When fitness landscapes are rugged, there is an intermediate regime where a multitude of emergent ecotypes---neither fully generalist nor fully specialist---may coexist.

Migration feedback produces an intriguing form of discontinuous transition which, although observed before \cite{brown1992evolution}, has not been characterized mathematically.
Of a distinct nature than previously studied catastrophic shifts \cite{villa2015eluding}, these transitions feature simultaneous stability swaps and reflect an underlying symmetry. 
Interestingly, when the symmetry of genotype-independent interactions is broken by a genotype-specific perturbation, the discontinuous transition splits into two closely-spaced continuous transitions (in analogy to how atomic spectral lines split due to spin-orbit coupling, which breaks spin-up/spin-down symmetry). This ``fine structure" occurs within simple ecological settings, suggesting that it could be observed in feasible experiments.

\section{Model}
Consider an asexual population inhabiting a network of $L$ discrete habitats connected by migration. Organisms are characterized by a genotype $i\in\{1,...,M\}$. Habitat $l\in\{1,...,L\}$ has a fitness landscape $\vec{\phi}_l\in\mathbb{R}^M$ where the $i$th component ($\phi_{l,i}$) specifies the fitness of genotype $i$ in habitat $l$. The population of habitat $l$ is denoted by $\vec n_l\in\mathbb{R}^M$ (component $n_{l,i}$ is the population of genotype $i$ in habitat $l$). Mutations occur at rate $\gamma$. Organisms migrate from habitat $l$ to $p$ at a per capita rate $k_{lp}$. Resources are limited, and each habitat has a resource limitation strength $\rho_l$ which sets the strength of a logistic discount to growth rates. We first consider a deterministic model of the population dynamics, and later investigate the effects of demographic noise. The deterministic model is the following system of coupled differential equations:
\begin{equation}\label{eom}
    \dot{\vec n}_l = (\hat V_l-\rho_l n_l^\textrm{tot}-k_l^\textrm{out})\vec n_l + \vec R_l
\end{equation}
for $l=1,...,L$, where the overdot indicates a time derivative and where $n_l^{\textrm{tot}}=\sum_{i=1}^{M}n_{l,i}$ is the total population of habitat $l$. Note that the logistic discount to net growth is identical for all genotypes. $k_l^\mathrm{out}$ is habitat $l$'s total out-going migration rate ($k_l^\mathrm{out}=\sum_{p=1}^L k_{lp}$) and $\vec R_l$ is the \textit{flux} of organisms migrating into habitat $l$ ($\vec R_l=\sum_{p=1}^L k_{pl}\vec n_p$). Fig. 1A is a schematic of a generic habitat $l$ with influx $\vec R_l$ and outflux $k_l^\mathrm{out}\vec n_l$. The `evolution matrix' $\hat V_l$ describes selection and mutation,
\begin{equation}\label{eq:V}
    \hat V_{l,ij} = \delta_{ij} \phi_{l,i} +\gamma \Big( \hat A_{ij} - \delta_{ij}\sum_{k=1}^M \hat A_{ki} \Big)
\end{equation}
where $\hat A$ is the genotype adjacency matrix: $\hat A_{ij}=1$ if genotypes $i$ and $j$ differ by a single mutation, and $\hat A_{ij}=0$ otherwise. Importantly, our results are valid for any symmetric $\hat A$. In the examples that we discuss, we choose $\hat A_{ij}=1$ if $|i-j|=1$ and zero otherwise (which is the adjancency matrix of a 1 dimensional lattice) because it makes the results easy to visualize. However, \textit{our results do not depend on this choice of $\hat A$}. This model extends that of \cite{waclaw2010dynamical,barkan2023multiple} to a network and, crucially, incorporates two-way migration between connected habitats.

\begin{figure}
\centering
\includegraphics[width=8.7cm]{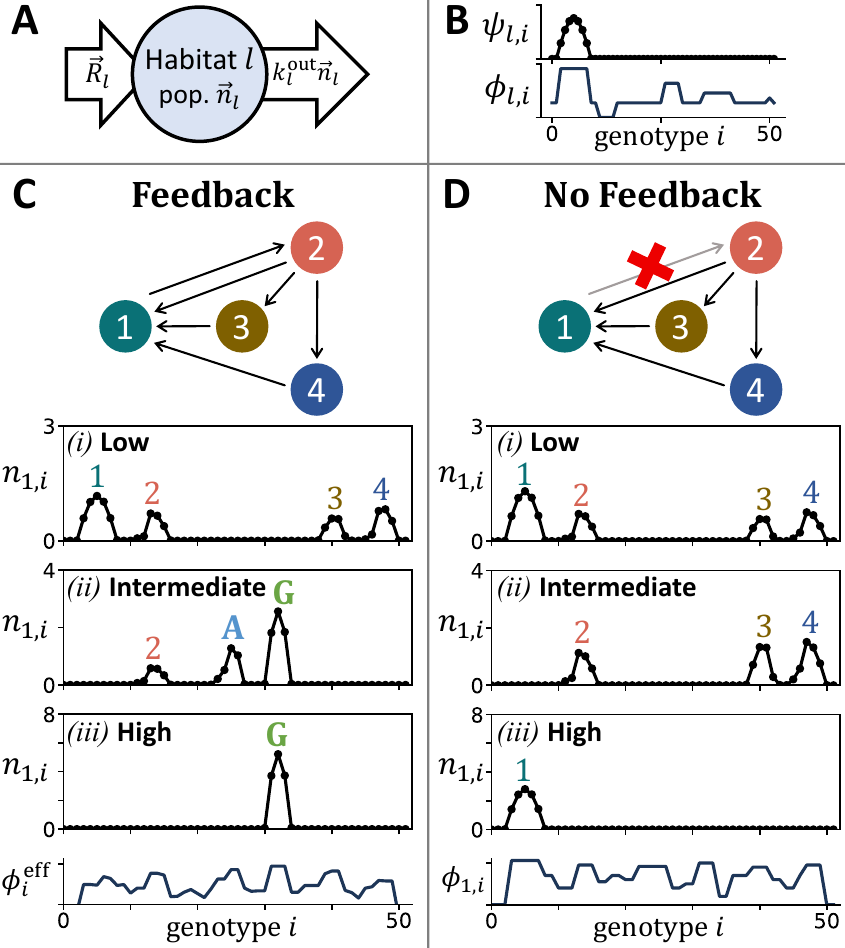}
\caption{\textbf{Migration feedback produces emergent ecotypes.}
\textbf{(A)} Schematic of a generic habitat $l$, indicating migration influx $\vec R_l$ and outflux $k_l^\mathrm{out}\vec n_l$.
\textbf{(B)} Example of a fitness landscape $\vec\phi_l$ and native ecotype $\vec\psi_l$, which is localized around the most fit region of the fitness landscape. 
\textbf{(C)} A network with feedback. Steady state populations of habitat 1 in the \textit{(i)} low, \textit{(ii)} intermediate, and \textit{(iii)} high migration regimes are shown. Native ecotypes are labeled with numbers corresponding to their habitat. Emergent ecotypes are labelled A and G. The effective landscape $\vec\phi^\mathrm{eff}$ in the high migration regime is shown, illustrating that the generalist G is the `native' ecotype on this effective landscape.
\textbf{(D)} A network with no feedback, obtained by removing one connection from the network in panel C. Without feedback, only native ecotypes can evolve. In the high migration regime, the effective landscape is merely the landscape of habitat 1, which is a sink.
The mutation rate is $\gamma=0.01\phi_\mathrm{max}$, where $\phi_\mathrm{max}$ is the maximum fitness value across all habitats. See SM for the steady state populations in other habitats and specification of other parameters.}
\label{fig:1}
\end{figure}

\section{Results}
\subsection{Evolution of native and emergent ecotypes}
The dynamics of Eq.~\ref{eom} lead to the evolution of ecotypes, which are subpopulations that form a cluster in genotype space as they specialize to a particular niche. Fig. 1B illustrates an ecotype in the simplest setting: an isolated habitat with zero migration. With the fitness landscape $\vec\phi_l$, the population evolves into a cluster with distribution $\vec\psi_l$ around the most-fit region of the fitness landscape (Fig. 1B). This cluster is the \textit{native ecotype} of the habitat. Mathematically, the steady state population is $\vec{n}^\mathrm{Nat}_l=K_l\vec{\psi_l}$ where $\vec\psi_l$ is the eigenvector of $\hat V_l$ with the largest eigenvalue ($\lambda_l$), normalized so that $\sum_i\psi_{l,i}=1$ \cite{waclaw2010dynamical,barkan2023multiple}. $K_l\equiv\lambda_l/\rho_l$ is the carrying capacity. The eigenvector $\vec\psi_l$ describes the distribution of genotypes in habitat $l$'s native ecotype, and it is exponentially localized around the region of genotype space most fit in habitat $l$ \cite{barkan2023multiple, waclaw2010dynamical}. 

In a network of habitats, the migration pattern $\{k_{lp}\}$ that connects the habitats has a profound effect on both the populations that evolve, and the response of evolved populations to changes in migration. In particular, networks with \textit{migration feedback} show emergent behaviors and abrupt critical transitions that are not exhibited by networks without feedback. To define this terminology: a network has \textit{migration feedback} if migration allows an organism to circulate back to a habitat it earlier left.
Definitions of this and related concepts are given in Appendix A. Fig. 1C shows a network with feedback (this network is also strongly connected; see Appendix A). In contrast, in a network without feedback, all migration leads eventually to \textit{sinks}, which are habitats with no outgoing flux. Fig. 1D shows a network without feedback, where habitat 1 is a sink.
 
Migration feedback allows for the evolution of \textit{emergent ecotypes}: ecotypes that are not native to any habitat. We demonstrate that, without feedback, only the native ecotypes of individual habitats can stably persist (Appendix B). In other words, while each habitat provides a local niche where its native ecotype can specialize, migration feedback generates non-local niches where emergent ecotypes are stabilized by the migration pattern through the network. Hence, according to the model, an experimental observation of emergent ecotypes implies that migration patterns must have feedback.

With or without migration feedback, network behaviors can be classified into three migration-dependent regimes. With feedback, there is \textit{(i)} a low migration regime with coexisting native ecotypes; \textit{(ii)} an intermediate migration regime in which one or more native ecotypes go extinct and are replaced by emergent ecotype(s). We only observe this regime when fitness landscapes are rugged, with multiple local fitness maxima. Finally, there is \textit{(iii)} a high migration regime in which all spatial coexistence of ecotypes is lost and an \textit{emergent generalist} takes over each strongly connected component of the network. In networks without feedback, no emergent ecotypes evolve, and the intermediate and high migration regimes are characterized by loss of native ecotypes without replacement by emergent ecotypes. The following two sections explain these regimes.

\subsubsection{Networks with feedback}
\textit{(i) Low migration regime: coexistence of native ecotypes.} For low rates, migration leads to spatial coexistence of multiple ecotypes within each habitat. We define a parameter $\kappa$ to control the overall strength of migration, and write the migration rates as $k_{lp}=\kappa r_{lp}$ where $r_{lp}$ specify the relative migration rates between pairs of habitats. For low $\kappa$, the steady-state populations are 
\begin{equation}\label{low_k}
    \vec{n}_l^* = \left(1-\kappa\chi\right)\vec{n}^\mathrm{Nat}_l + \kappa \sum_{p\neq l}r_{pl}\hat F_l\vec{n}_p^\mathrm{Nat} + \mathcal{O}(\kappa^2)
\end{equation}
where $\hat F_l$ (defined in SM) is a linear filter that describes how the genotype distribution of an ecotype migrating into habitat $l$ is modified by the local selection pressure. $\chi$ is a constant (defined in SM). Eq. \ref{low_k} describes how the native ecotype in each habitat coexists with the (filtered) native ecotypes of other habitats. This behavior is illustrated in Fig. 1C\textit{(i)}, which shows how habitat 1 supports its native ecotype as well as incoming ecotypes from all other habitats (see SM for the steady state populations in other habitats).

\textit{(ii) Intermediate migration regime: growth of emergent ecotypes.} As $\kappa$ increases, the average fitness of native ecotypes decreases because migration carries native ecotypes away from the habitats where they are most fit. At the same time, the average fitness rises for other genotypes with moderate fitness across multiple habitats connected by feedback. This intuitive idea is made more precise in Appendix B. This results in cluster(s) of globally-fit genotypes (i.e. emergent ecotypes) growing  and outcompeting native ecotype(s). These emergent ecotypes persist while one or more native ecotype goes extinct. This behavior is illustrated in Fig. 1C\textit{(ii)}, where two emergent ecotypes (labelled A and G) coexist and three native ecotypes (1, 3 and 4) have gone extinct. This regime typically contains many phases, each characterized by a particular set of coexisting native and emergent ecotypes.

\textit{(iii) High migration regime: dominance of an emergent generalist.} For very high migration rates, the dynamics dramatically simplify because the timescale on which the population samples the network becomes fast relative to net population growth. In the limit $\kappa\to\infty$, each strongly connected component of the network behaves as a single isolated habitat; it has a population $\vec N=\sum_l \vec n_l$ (summing over habitats within the component) governed by dynamics of the form of Eq.~\ref{eom} but with effective parameters:
\begin{equation}\label{eom_eff}
    \dot{\vec{N}} = (\hat V^\textrm{eff}-\rho^\textrm{eff} N^\textrm{tot})\vec N
\end{equation}
where $\hat V^\textrm{eff}_{ij}=\delta_{ij}\phi^\textrm{eff}_i+\gamma(A_{ij} - \delta_{ij}\sum_k A_{ki})$ with $\vec\phi^\textrm{eff}=\sum_l s_l \vec\phi_l$ and $\rho^\textrm{eff}=\sum_l s_l^2\rho_l$. The weights $s_l$ are the stationary probabilities for a random walk on the component with hopping rates $r_{lp}$ (see SM). Eq. \ref{eom_eff} reveals that, at high migration rates, only one ecotype can stably persist within each strongly connected component: the `native' ecotype of the effective fitness landscape $\vec\phi^\textrm{eff}$. We call this ecotype the \textit{emergent generalist} and it has population $\vec{N}^*=K_\mathrm{eff}\vec{\psi}_\mathrm{eff}$ where the carrying capacity $K_\mathrm{eff}$ and genotype distribution $\vec{\psi}_\mathrm{eff}$ are defined as before, but now with the effective habitat parameters.

Even though Eq.~\ref{eom_eff} is only exact in the $\kappa\to\infty$ limit, the generalist typically emerges at migration rates much lower than the maximum per capita growth rate in the system (see SM). Hence, the high-$\kappa$ regime is likely accessible in both laboratory and natural systems. Fig.~\ref{fig:1}C\textit{(iii)} shows the emergent generalist (labeled $G$) and the effective fitness landscape $\vec\phi^\mathrm{eff}$ for the example network shown.

\subsubsection{Networks without feedback}
In networks with no feedback, there is an analytical formula for the steady state populations at any $\kappa$ (see SM). This formula shows that only native ecotypes can stably exist. There is \textit{(i)} a low migration regime with the same behavior as for networks with feedback (described by Eq.~\ref{low_k}). As $\kappa$ increases into the \textit{(ii)} intermediate regime, a native ecotype will go extinct; however, it is possible that this ecotype will regain stability at even higher migration rates through a re-entrant transition (see \cite{barkan2023multiple}). For higher $\kappa$, there can be a variety of phases where various native ecotypes are extinct. For sufficiently large $\kappa$ in the \textit{(iii)} high migration regime, the populations will be depleted due to outgoing migration in every habitat $l$ with $k^\textrm{out}_l\neq 0$. Only \textit{sink} habitats will have nonzero population, and only the native ecotype in each sink habitat will persist. Hence, similar to networks with feedback, all spatial coexistence is lost in the high migration regime, but unlike networks with feedback, there is no generalist adaptation. Notably, without feedback, much higher migration rates are typically required for the system to enter the high migration regime (see SM). Fig. 1D illustrates three regimes for a network with no feedback.

\subsection{Critical transitions and phase diagram}

Interestingly, migration feedback not only allows a greater variety of ecotypes to evolve, it also induces a richer variety of critical transitions. These transitions occur at migration rates where an ecotype gains or loses stability (i.e. where an ecotype's steady state population transitions from zero to nonzero, or vice versa). It is illuminating to explore these phenomena within a minimal network: two interconnected habitats, as shown in Fig. 2A. In this minimal setting, we map out the phase diagram and examine the critical transitions that separate the phases. The two-habitat system has migration feedback when both $k_{12}$ and $k_{21}$ are positive.

\begin{figure*}
\centering
\includegraphics[width=17.8cm]{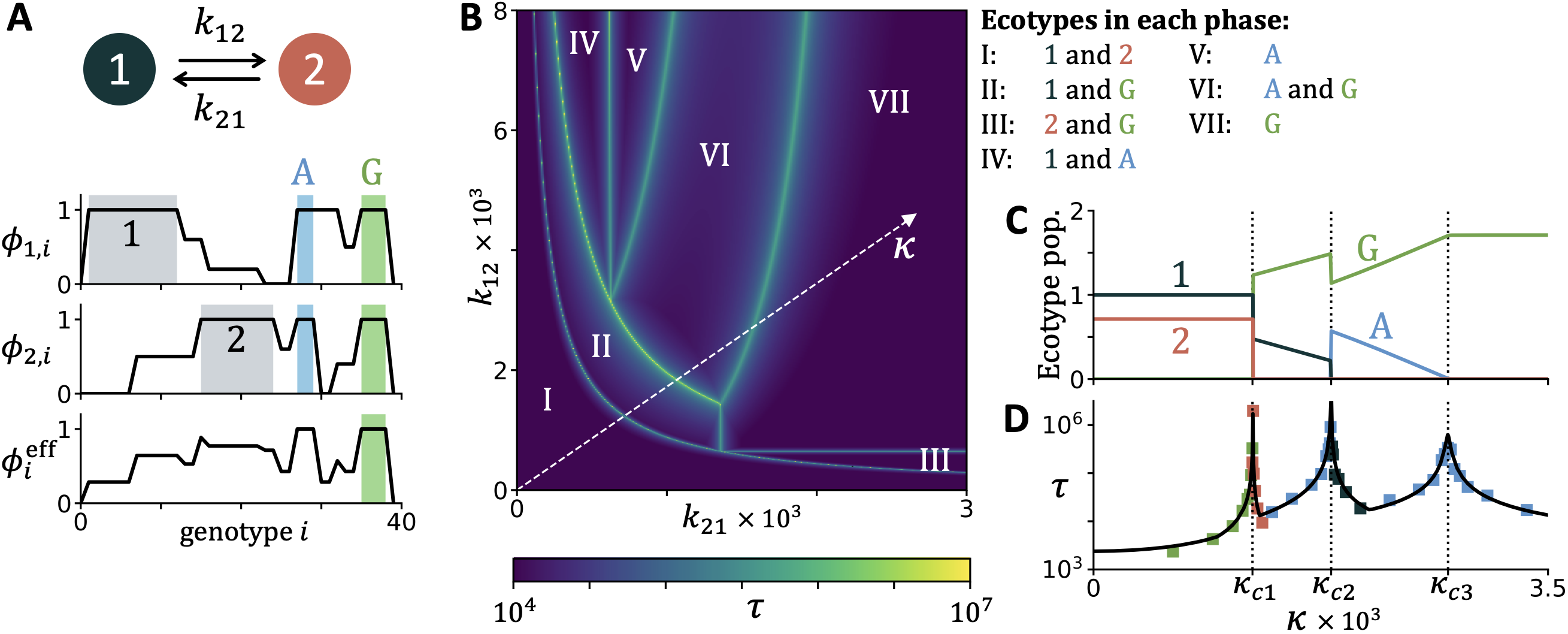}
\caption{\textbf{Continuous and discontinuous critical transitions in a minimal setting.} \textbf{(A)} Two-habitat system with fitness landscapes $\vec\phi_1$ and $\vec\phi_2$. Shaded regions indicate the parts of genotype space corresponding to native ecotypes 1 and 2 and emergent ecotypes $A$ and $G$. Effective landscape $\vec\phi^\mathrm{eff}$ in the high $\kappa$ regime where $r_{12}=1$ and $r_{21}=0.6$. These fitness landscapes were chosen for visual clarity, though all conclusions apply generally, as supported by both analytical results and statistical tests with randomly generated fitness landscapes (see SM). \textbf{(B)} Phase diagram revealed by a heat map of the relaxation time $\tau$. Phases (labelled by Roman numerals) are characterized by the ecotypes that stably coexist; the coexisting ecotypes within each phase are listed. \textbf{(C)} Steady state total populations of each ecotype (summed over both habitats) as a function of migration strength $\kappa$, along a slice of the phase diagram indicated by the dotted white line in panel B. \textbf{(D)} Relaxation time $\tau$ (solid black curve) along the same slice as panel C. $\tau$ is peaked at both the continuous ($\kappa=\kappa_{c3}$) and discontinuous ($\kappa=\kappa_{c1}, \kappa_{c2}$) phase transitions. Colored squares show the autocorrelation time of the population of the ecotype that emerges or vanishes at the nearby transition, obtained from stochastic simulations with demographic noise (see SM). Color coding is as labelled in panel C. From left to right, the ecotypes whose autocorrelation times are shown are: G, 2, A, 1, A. Autocorrelation times for all ecotypes at each $\kappa$ are provided in the SM.}\label{fig:2}
\end{figure*}

Fig.~\ref{fig:2}A shows example fitness landscapes, and the effective landscape in the high migration regime for a particular ratio of $k_{12}/k_{21}$ (the slope of the dotted line in Fig. 2B). The shaded regions indicate the ecotypes that can exist in this system. The regions labelled 1 and 2 correspond to the native ecotypes of habitats 1 and 2, and, as we will show, there are two emergent ecotypes A and G, as indicated.

Fig. 2B shows the phase diagram for this system. Each phase (labelled I through VII) is characterized by the ecotypes that stably exist. Each point $(k_{21},k_{12})$ on the diagram is colored according to the relaxation time $\tau$ of the population at the migration rates $k_{21}$ and $k_{12}$. This relaxation time to steady state is given by $\tau=-1/w_1$ where $w_1$ is the greatest (i.e. least negative) eigenvalue of the Jacobian matrix of the system defined by Eq.~\ref{eom}. $\tau$ is sharply peaked around the migration rates at which an ecotype gains or loses stability due to the phenomenon of critical slowing down (CSD). Mathematically, CSD occurs due to a bifurcation in the steady state at the critical transition; as we will see, feedback induces an unusual form of bifurcation. Thus, the bright lines in Fig. 2B along which $\tau$ is sharply peaked indicate where the critical transitions occur.

The phase diagram provides a global view of the low, intermediate, and high migration regimes. Phase I is the low migration regime, phase VII is in the high migration regime, and the remaining phases constitute the intermediate migration regime. There are three other phases in the high migration regime (out of view, shown in SM) in which different $k_{12}/k_{21}$ ratios produce different $\vec \phi^\mathrm{eff}$ with corresponding emergent generalists. 

Fig. 2C shows one `slice' through the phase diagram along the white dotted line in Fig. 2B. This reveals two distinct types of critical transitions. There are \textit{discontinuous transitions}, where the steady state populations change discontinuously across the transition, at $\kappa=\kappa_{c1}$ and $\kappa_{c2}$. At these transitions a native ecotype is replaced by an emergent ecotype. There is a \textit{continuous transition} (where the steady state populations change continuously) at $\kappa=\kappa_{c3}$, where an ecotype goes extinct without any replacement by another ecotype. It is no coincidence that the transitions where an emergent ecotype replaces a native ecotype are discontinuous, whereas transitions involving only the extinction of an ecotype are continuous. Moreover, when feedback is removed ($k_{12}=0$ or $k_{21}=0$), no discontinuous transitions occur (see SM). 

These observations point toward a general connection between migration feedback, emergent ecotypes, and discontinuous transitions. First, emergent ecotypes require migration feedback to exist (as discussed above). Second, discontinuous transitions require emergent ecotypes. Indeed, we show in Appendix B that a transition at which a native ecotype is replaced by an emergent ecotype is always discontinuous. Conversely, in networks without feedback, transitions cannot be discontinuous; instead, only continuous transitions due to a single transcritical bifurcation can occur \cite{barkan2023multiple}. These results show that feedback induces a greater variety of emergent phenomena, and that observation of emergent ecotypes or discontinuous transitions in a particular system implies that migration patterns must contain feedback.

As $\kappa$ approaches a discontinuous transition, there is no change in the steady state populations to warn of the impending transition. However, CSD may provide an early warning signal. In an experiment, measuring $\tau$ directly requires measuring the dynamics of each genotype, which may be unfeasible. 
Instead, one needs only to measure ecotype populations and estimate $\tau$ from the autocorrelation time of stochastic fluctuations; see SM for details. Fig. 2D shows both $\tau$ (black curve) and the autocorrelation times of the ecotype that emerges (or vanishes) at each transition (squares, colored according to ecotype). As shown, a rising $\tau$ indicates an approaching discontinuous transition even though steady state populations show no decline prior to the transition. An important caveat is that, in order for this warning signal to work, the genotypes that comprise the emerging but not-yet-stable ecotype must be present at low levels in the system. This may occur in the human gut, where a broad range of microbes enter from food, so that there are low levels of many genotypes that are otherwise unable to persist. It could also occur when mutation rates are sufficiently high, possibly in rapidly evolving B-cell populations. Moreover, in synthetic experiments, organisms could be introduced by design to satisfy this requirement. This caveat is further discussed in the SM.

Questions remain: What causes the discontinuity in steady state populations? Are there feasible experiments that could probe this abrupt change? These questions are addressed in the following section.

\subsection*{Mechanism and fine-structure of the discontinuous transitions}

The mechanism of the discontinuous transitions can be elucidated by reducing Eq.~\ref{eom} to a simple and easily solvable model. As we show below, the discontinuity arises from two transcritical bifurcations that occur at the same value of $\kappa$, reflecting symmetry in the governing equation. A genotype-dependent perturbation that breaks the symmetry of genotype-independent interactions splits these simultaneous bifurcations into two nearby ones, inducing a ``fine structure" to the transition.

Specifically, the ecological character of the transition, where ecotype emergence or extinction involves an entire cluster of genotypes collectively emerging or vanishing, motivates a simple model of ecotype population dynamics. Consider a two-habitat system with three ecotypes: native ecotypes of habitats 1 and 2 and an emergent generalist $G$. In general, the populations $n_{le}$ ($l\in\{1,2\}$, $e\in\{1,2,G\}$) obey Eq.~\ref{eco_model} below, but to retain only the essential mechanism, we further reduce the model to two degrees of freedom. By assuming that the native ecotypes both have fitness $\phi$ in their native habitat and that the generalist has fitness $\mu$ in both habitats, and assuming $\rho_1=\rho_2=\rho$ and $k_{12}=k_{21}=\kappa$, the model reduces to (see SM):
\begin{align}
    \dot x_N &= (\phi-\rho(x_N+x_G)-\kappa)x_N \label{simple1}\\
    \dot x_G &= (\mu-\rho(x_N+x_G))x_G \label{simple2}
\end{align}
where $x_N$ is the population of the native ecotype in each habitat and $x_G$ the population of the generalist in either habitat. This model has two steady states that swap stability at a discontinuous transition at critical migration $\kappa_c=\phi-\mu$ (see SM). Fig. 3A shows the stable (solid lines) and unstable (dashed lines) steady state values of $x_N$, and the discontinuous transition at $\kappa_c$. While a transition of this type has been described before \cite{brown1992evolution}, its exact nature has not been clarified.

\begin{figure}
\centering
\includegraphics[width=8.7cm]{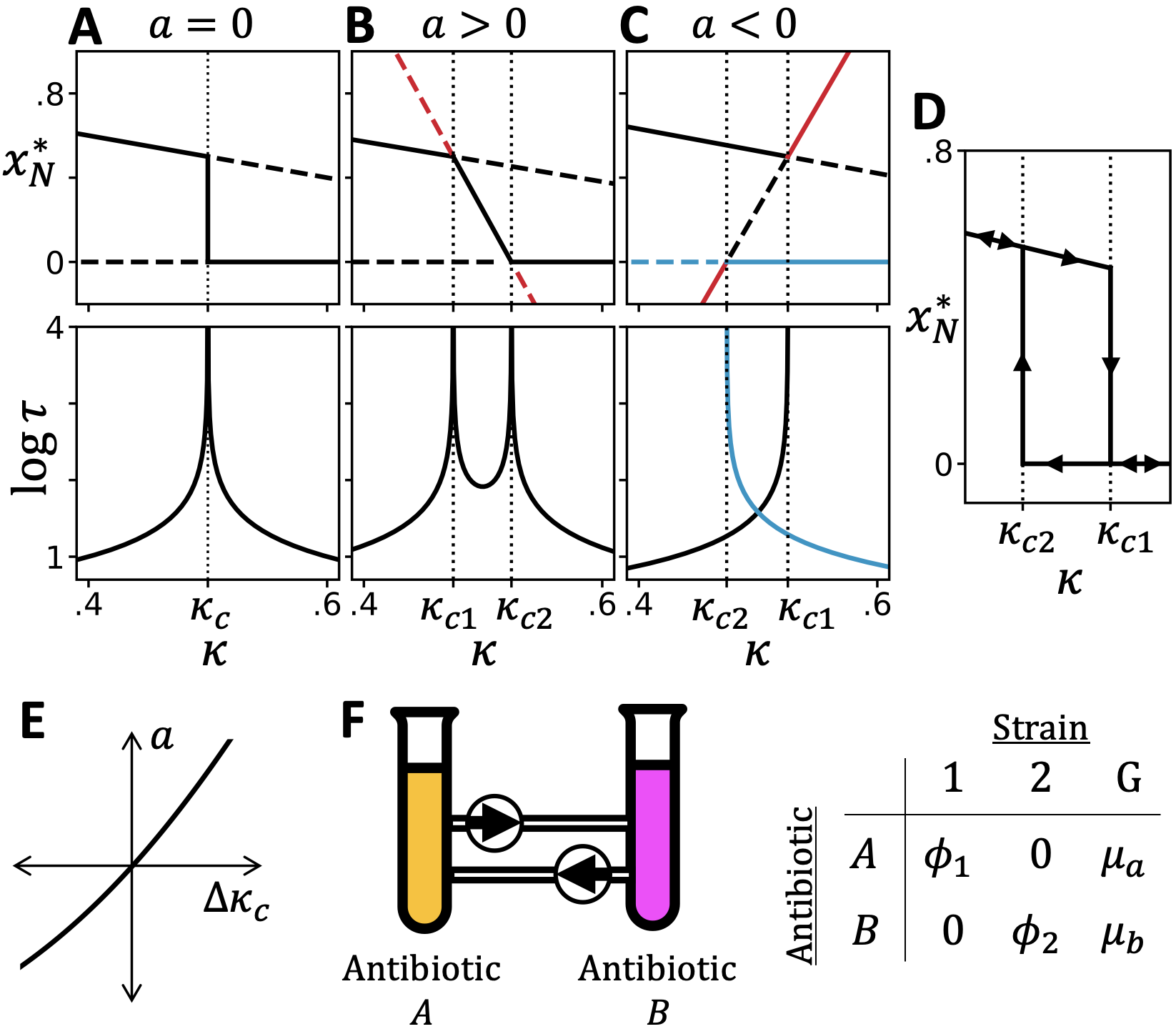}
\caption{\textbf{Fine-structure of discontinuous transitions permits experimental probing for asymmetry of interactions.} \textbf{(A)} The steady state values $x^*_N$ when $a=0$, showing a discontinuous transition. Stable and unstable steady states are shown as solid and dashed lines, respectively. Lower panel shows the relaxation time $\tau$. See SM for $x^*_G$ for panels A, B, and C. \textbf{(B)} $x^*_N$ for $a>0$, where the discontinuous transition has split into two continuous transitions. Each transition is a transcritical bifurcation, where two steady states swap stability (i.e. switch from solid to dashed or vice versa). Red dashed lines indicate unrealistic steady states with negative populations ($x_N^*<0$ or $x_G^*<0$). Lower panel shows that $\tau$ is peaked around both transitions. \textbf{(C)} $x^*_N$ for $a<0$, where there is a region of bistability between the two continuous transitions. Red lines indicate steady states with negative populations. Lower panel shows $\tau$ at each stable steady state, which shows two one-sided peaks. The stable steady state $x_N^*=0$ and its corresponding relaxation time $\tau$ are marked in blue, to identify which $\tau$ corresponds to which $x_N^*$. \textbf{(D)} For $a<0$, sweeping $\kappa$ across both transitions and back will produce a hysteresis loop as shown. \textbf{(E)} Relation between $\Delta\kappa_c=\kappa_{c2}-\kappa_{c1}$, which could be measured experimentally, and strength $a$ of asymmetry of interactions. \textbf{(F)} Schematic of an experiment involving three antibiotic resistant bacterial strains -- two specialists and one generalist. Migration is controlled by transporting bacteria between two habitats (e.g. chemostats). The table lists the growth rate of each strain in the habitat with either antibiotic. }
\label{fig:3}
\end{figure}

Intriguingly, at the transition, the restoring force that drives the system to steady state vanishes \textit{exactly} (in a typical bifurcation, it would vanish only up to certain order). To be specific, for a population $[x_N\; x_G]=[x_N^*\; x_G^*]+c(t)\bm\zeta$ where $[x_N^*\; x_G^*]$ is a steady state and $\bm\zeta$ the eigenvector of the Jacobian with zero eigenvalue, the dynamics are $\dot c = 0$. To explore its implications, we perturb Eqs.~\ref{simple1} and \ref{simple2} with the lowest-order interaction terms $-a\rho x_N^2$ and $-a\rho x_G^2$ that break the symmetry of ecotype-independent interactions. These terms reintroduce a restoring force at the transition, and by taking the limit $a\to0$ we reveal how a lack of restoring force induces a discontinuity.

The perturbation introduced by the above symmetry-breaking terms splits the discontinuous transition into two closely-spaced transcritical bifurcations: one at $\kappa_{c1}=\phi-\mu(1+a)$ and the other at $\kappa_{c2}=\phi-\mu/(1+a)$. The resulting steady state populations are shown in Fig. 3B (for $a>0$) and Fig. 3C (for $a<0$). At each bifurcation, a steady state (solid line) swaps stability with an unstable steady state (dashed line). Red lines mark unrealistic steady states (with negative populations). For $a<0$, the region between $\kappa_{c2}$ and $\kappa_{c1}$ exhibits bistability, where an external input (drive or noise) can induce a switch from one steady state to the other. Intuitively, this bistability arises because cooperativity ($a<0$) stabilizes ecotypes with large populations, so the system is stable either when ecotype 1 has a high population and ecotype 2 has zero population, or vice versa. As a result, the system will show hysteresis if $\kappa$ drifts across both transitions and back (Fig. 3D). This would provide an experimental signature that distinguishes between competitive ($a>0$) and cooperative ($a<0$) within-ecotype interactions. Furthermore, the strength $a$ of the cooperative or competitive effect could be determined experimentally by measuring the split between transitions $\Delta\kappa_c\equiv\kappa_{c2}-\kappa_{c1}$ (Fig. 3E).

An experiment that not only detects the fine-structure of a discontinuous transition, but also uses it as a \textit{tool} to measure inter-organism interactions, could be arranged as follows. Take a bacterial species with three strains: $1$, $2$, and $G$. Strains $1$ and $2$ are each resistant, respectively, to the antibiotics $A$ and $B$. Strain $G$ is resistant to both antibiotics, but faces a fitness cost due to the maintenance of resistance. Habitats 1 and 2 (implemented, for example, with two chemostats) contain antibiotic A and B, respectively. A cartoon of an experimental setup is shown in Fig. 3F, where bacteria are transported between the habitats at a controlled rate. 
The table in Fig. 3F shows the growth rate of each strain in either habitat, where $\phi_1>\mu_a$ and $\phi_2>\mu_b$. In the special case that $\phi_1=\phi_2=\phi$ and $\mu_1=\mu_2=\mu$, the dynamics reduce to Eqs.~\ref{simple1} and \ref{simple2}. In the more general case where these conditions do not hold, both a continuous and a discontinuous phase transition are predicted (see SM). The fine-structure of the discontinuous transition would reveal whether intra-strain interactions are more or less competitive than inter-strain interactions (Fig. 3E). Moreover, one could use this setup to quantify the effects of interaction-altering mutations.

\section*{Discussion}

Spatial heterogeneity is a nearly universal feature of the environments in which biological populations live and evolve, and the impacts of heterogeneity on evolutionary and ecological dynamics have been heavily studied \cite{felsenstein1976theoretical,chesson2000mechanisms, amarasekare2003competitive,dias1996sources, kawecki2004conceptual,hermsen2012rapidity, rainey1998adaptive, pearce2020stabilization,roy2020complex,hu2022emergent,lombardo2014nonmonotonic,barkan2023multiple}. Yet, much remains to be understood about how differing migration patterns affect such systems. We show a fundamental connection between migration patterns, the populations that evolve, and the response of those populations to changes in migration. Specifically, migration feedback is required for emergent ecotypes and generalists to evolve, and the emergence of such ecotypes occurs through an intriguing form of discontinuous critical transition. These abrupt transitions occur only in systems with migration feedback. The significance of feedback in a network of habitats is reminiscent of the significance of feedback loops (i.e. `recurrence') in recurrent neural networks (RNNs). RNNs allow for memory (persistence of information in time) \cite{lipton2015critical}, which resembles how migration feedback generates non-local niches (persistence of fitness information through spatially-extended collections of habitats).

Since migration feedback (circulating flux) can occur in a multitude of evolving systems throughout the body, the pertinent emergent ecotypes may have important health implications. During metastatic invasion, cancer cells not only move out toward distant ecological niches, but also can return to the primary tumor \cite{norton1988gompertzian}. Our model suggests that under high migration (relative to net growth), drug-resistant cancer cells may evolve as a generalist ecotype that fills a subnetwork of connected tissue sites. This might provide an alternative explanation for cancer recurrence (regrowth after a surgery): rather than due to reseeding, recurrence in time is in fact recurrence in space; after a tumor is removed from one location, generalist cancer cells may continue to exist elsewhere, and eventually spread back to the original location. 
On the other hand, bacteria adapt in habitats connected by circulating fluid flow (e.g. in the gut and lungs), and the evolution of antibiotic resistance proceeds on rugged fitness landscapes \cite{palmer2015delayed, papkou2023rugged}. Our model predicts that there is an intermediate migration regime in which, by modulating their own motility, bacteria can exploit a variety of emergent ecotypes for adapting in a broader range of niches.  

Critical transitions and emergent phenomena have been widely studied in ecology \cite{wissel1984universal, scheffer2001catastrophic, van2007slow,drake2010early,dai2012generic,scheffer2015generic,villa2015eluding,dakos2019ecosystem,xu2023non,bunin2017ecological,opper1992phase,hu2022emergent}. A form of discontinuous transition was recently discussed in \cite{villa2015eluding} where an abrupt change occurs due to loss of bistability. The discontinuous transitions in our model are different, as they do not require bistability and arise due to simultaneous transcritical bifurcations that reflect an underlying symmetry (see SM). 
One biological consequence our model predicts is that strengthening migration causes a decline in species abundance as an emergent ecotype drives native ecotypes to extinction. This resembles what is found experimentally \cite{hu2022emergent} that species abundance declines in response to strengthening interactions.


Feasible experiments may be able to explore the predictions of our model. We propose such an experiment in which only three bacterial strains and two habitats are required. Measurement of the fine-structure of the transition, where genotype-dependent interactions split the discontinuous transition into two closely-spaced continuous transitions, may provide a means to infer the type and magnitude of asymmetry in interactions. Alternatively, interactions may be modified experimentally (e.g. by varying nutrient concentrations as in \cite{ratzke2020strength, hu2022emergent}) in order to tune the fine structure. Finally, expanding our model to explicitly include resource dynamics may reveal additional behaviors, such as multi-stability (such as in \cite{dubinkina2019multistability}), that such experiments could detect.

\appendix

\section{Definition of migration feedback}

We introduce the terminology of \textit{migration feedback}, which is similar to, but not equivalent to, the property of \textit{strong connectivity} in graph theory. Here are our definitions:

\begin{enumerate}
    \item A network of habitats has \textit{migration feedback} if there is (at least one) habitat that an organism can migrate out of, then return to at a later time. Equivalently, a network has migration feedback if there is a pair of strongly connected habitats.
    \item A network is \textit{strongly connected} if any node (i.e. habitat) can be reached from any other node through some sequence of steps. Note that all strongly connected networks of habitats have feedback, but not all networks with feedback are strongly connected.
    \item A \textit{strongly connected component} of a network is a subnetwork that is strongly connected, and for which the inclusion of any additional nodes in the subnetwork would break its strong connectivity (in other words, the subnetwork is as large as possible while maintaining its strong connectivity).
    Under high migration, each strongly connected component is overtaken by a single emergent generalist.  
\end{enumerate}

\section{Feedback, emergent ecotypes, and discontinuous transitions}

Migration feedback, emergent ecotypes, and discontinuous transitions are connected in the following way: emergent ecotypes require migration feedback to exist at steady state, and discontinuous transitions require an emergent ecotype to occur. Hence, migration feedback is a necessary ingredient for both of these behaviors.

To prove that feedback is required for emergent ecotypes and discontinuous transitions, we prove that networks without feedback only produce native ecotypes and continuous transitions. The proof of this follows from the analytical formula for steady state populations in networks without feedback, and it is given in the SM.

The following illustrates intuitively why networks with feedback produce emergent ecotypes and discontinuous transitions. Feedback allows global growth rates to be larger than \textit{relative local growth rates} (defined below), which allows emergent ecotypes to grow on the network despite having lower fitness in each habitat than the habitat's native ecotype. Second, the discontinuity in steady state populations at a discontinuous transition requires that the ecotype transitioning from unstable to stable has an intermediate fitness (not the maximum or minimum fitness in any habitat). This condition is satisfied only for emergent ecotypes.

Consider the transition from the low migration regime to the intermediate regime, where the first emergent ecotype arises and coexists with the native ecotypes. We will model this using an ecological model (neglecting mutation) of the form
\begin{equation}\label{eco_model}
    \dot n_{le} = (\phi_{le}-\rho_l n_l^\mathrm{tot}-k_l^\mathrm{out})n_{le} + \sum_p k_{pl}n_{pe}
\end{equation}
where subscript $l$ denotes the habitat ($l=1,...,L$), and subscript $e$ denotes the ecotype ($e=1,...,L$ are the native ecotypes and $e=G$ is a non-native ecotype which emerges when the system enters the intermediate migration regime). The connection between migration feedback, emergent ecotypes, and discontinuous transitions becomes clearer when the dynamics are expressed in terms of ecotype frequencies $f_{le}\equiv n_{le}/n_l^\mathrm{tot}$. In these new variables, the dynamics are
\begin{align}
    \dot f_{le} &= (\phi_{le}-\phi_l^\mathrm{avg})f_{le} + \sum_p k_{pl}\frac{n_p^\mathrm{tot}}{n_l^\mathrm{tot}}(f_{pe}-f_{le}) \label{fdot} \\
    \dot n_l^\mathrm{tot} &= (\phi_l^\mathrm{avg}-\rho_ln_l^\mathrm{tot}-k_l^\mathrm{out})n_l^\mathrm{tot} + \sum_p k_{pl}n_p^\mathrm{tot} \label{ntotdot}
\end{align}
where $\phi_l^\mathrm{avg}=\sum_e f_{le}\phi_{le}$. Now, consider the growth of the emergent ecotype $G$. Let $\vec f_G\equiv [f_{1G}\;...\;f_{LG}]^T$, so $\dot {\vec f}_G = M_G\vec f_G$ where
\begin{equation}
    M_G =
    \begin{bmatrix}
    M_{G,11} & k_{21}\frac{n_2^\mathrm{tot}}{n_1^\mathrm{tot}} & \hdots \\
    k_{12}\frac{n_1^\mathrm{tot}}{n_2^\mathrm{tot}} & M_{G,22} & \\
    \vdots & & \ddots
    \end{bmatrix}
\end{equation}
and where the diagonal elements
\begin{equation}
    M_{G,ll} =  \phi_{lG}-\left(\phi_l^\mathrm{avg}+\sum_p k_{pl}\frac{n_p^\mathrm{tot}}{n_l^\mathrm{tot}}\right)
\end{equation}
are the \textit{relative local growth rates} (i.e. the fitness of $G$ in habitat $l$, relative to average fitness in $l$ plus the discount due to migration out of $l$). This contrasts with the \textit{global} growth rates of $G$, which are the eigenvalues of $M_G$ (if an eigenvalue of $M_G$ is positive, then $G$ will grow on the network).

The key effect of feedback is that it allows global growth rates to be larger than relative local growth rates. The reason is that, without feedback, $M_G$ can be written in lower triangular form (with an appropriate labeling of habitats), whereas with feedback, $M_G$ cannot be written in lower triangular form. For a lower triangular matrix (no feedback), the diagonal elements are equal to the eigenvalues; in other words, relative local growth rates are equal to global growth rates. For a non-lower triangular matrix (with feedback), the eigenvalues can be larger than the diagonal elements, so feedback allows global growth rates to be larger than the relative local growth rates. 

Without feedback, emergent ecotypes always have negative relative local growth rates at steady state. Hence, an emergent ecotype can never grow without feedback. With feedback, instead, an emergent ecotype can have a positive global growth rate even if all relative local growth rates are negative.

To see why every discontinuous transition must involve the emergence of an emergent ecotype, it is useful to consider the discontinuous transition as the limit where two transcritical bifurcations occur simultaneously. First, consider a discontinuous transition that is split into two transcritical bifurcations at $\kappa_{c1}$ and $\kappa_{c2}$ by a symmetry-breaking interaction term (as in Fig.~3B). As shown in Fig. 3B, the steady state populations change rapidly as $\kappa$ is increased from $\kappa_{c1}$ to $\kappa_{c2}$. As the symmetry-breaking term is reduced toward zero, $\kappa_{c1}$ and $\kappa_{c2}$ move closer together, and the steady state $\bm n^*(\kappa)$ for $\kappa_{c1}<\kappa<\kappa_{c2}$ becomes `steeper' (i.e. $|\frac{d}{d\kappa}\bm n^*(\kappa)|$ becomes larger). $\bm n^*(\kappa)$ becomes `vertical' (varies arbitrarily quickly with $\kappa$) in the limit that the symmetry-breaking term vanishes (as in Fig. 3A). This `vertical' line of states represents a continuum of steady states at $\kappa_c$, parametrized by $\bm n^*+ c\bm\zeta$ for any $c$, where the steady state $\bm n^*$ and zero mode $\bm\zeta$ are vectors with components for each ecotype in each habitat.
In the language of dynamical systems theory, this continuum of states is the \textit{center manifold}, and at a discontinuous transition, the dynamics on the center manifold vanish to \textit{all} orders in $c$. Inserting $\bm n^*+ c\bm\zeta$ into Eq.~\ref{eom} shows that, in order for $\dot c=0$ for any $c$, $n_l^\mathrm{tot}$ must be independent of $c$ in every habitat. 

Now consider Eq.~\ref{ntotdot} for this continuum of steady states. $\dot n_l^\mathrm{tot}=0$, as required for a steady state, and the previous paragraph shows that $n_l^\mathrm{tot}$ is constant on this continuum of steady states. Therefore, Eq.~\ref{ntotdot} implies that $\phi_l^\mathrm{avg}$ must also be constant on this continuum of steady states. For $\phi_l^\mathrm{avg}$ to remain constant while $f_{lG}$ transitions from zero to nonzero, the condition $\min_e\{\phi_{le}\}<\phi_{lG}<\max_e\{\phi_{le}\}$ must be satisfied for every habitat $l$ (if this weren't true, then increasing $f_{lG}$ would necessarily increase or decrease $\phi_l^\mathrm{avg}$). Native ecotypes do not satisfy this inequality, because for native ecotype $l$, $\phi_{ll}=\max_e\{\phi_{le}\}$. Therefore, ecotype $G$ cannot be native to any habitat, so it must be an emergent ecotype.

\bibliography{main}

\end{document}